\documentclass[runningheads]{llncs}
\usepackage[T1]{fontenc}
\usepackage{graphicx}

\usepackage{hyperref}
\usepackage{color}

\usepackage{amsmath,amssymb,amsfonts}

\usepackage{algorithm}
\usepackage{algorithmicx}
\usepackage{algpseudocode}
\usepackage{xspace}
\usepackage{todonotes}
\usepackage{url}
\usepackage{enumitem}


\usepackage{graphicx}
\usepackage{pgfplots}
\usepackage{tikz}
\usetikzlibrary{automata, positioning}
\usetikzlibrary{shapes.geometric, arrows}

\usetikzlibrary{positioning, arrows.meta}
\usepackage{textcomp}
\usepackage{xcolor}

\definecolor{lightblue}{RGB}{173, 216, 230}

\usepackage{subcaption}
\pgfplotsset{compat=1.17} 

\usepackage{bbding}
\usepackage{biblatex}
\definecolor{darkgreen}{rgb}{0.0, 0.35, 0.0} 

\def\BibTeX{{\rm B\kern-.05em{\sc i\kern-.025em b}\kern-.08em
    T\kern-.1667em\lower.7ex\hbox{E}\kern-.125emX}}

\newcommand{\cmnt}[1]{}
\newcommand{\ignore}[1]{}
\newcommand{\remove}[1]{}

\newcommand{\slsqp} {SLSQP\xspace}
\newcommand{\qrf} {QR Factorization\xspace}

\begin{document}

\title{Efficient Task Graph Scheduling for Parallel QR Factorization in SLSQP\thanks{GitHub Repository: \url{https://github.com/PDCRL/ParSQP}}}
%
%
\author{
Soumyajit Chatterjee \inst{1} \Envelope  \and
Rahul Utkoor\inst{2}\and
Uppu Eshwar\inst{1} \and
Sathya Peri\inst{1}\and
V.Krishna Nandivada\inst{3}
}

\institute{Indian Institute of Technology, Hyderabad\\
\email{\{ai22mtech02005@, ch21btech11034@, sathya\_p@cse}\}.iith.ac.in \and
QUALCOMM India Private Limited\\
\email{rutkoor@qti.qualcomm.com} \and
Indian Institute of Technology, Madras\\
\email{nvk@iitm.ac.in}
}

\titlerunning{Efficient Task Graph Scheduling for Parallel QR Factorization}
\authorrunning{Chatterjee et al.}

\maketitle              

\begin{abstract}
Efficient task scheduling is paramount in parallel programming on multi-core architectures, where tasks are fundamental computational units. QR factorization is a critical sub-routine in Sequential Least Squares Quadratic Programming (SLSQP) for solving non-linear programming (NLP) problems. QR factorization decomposes a matrix into an orthogonal matrix Q and an upper triangular matrix R, which are essential for solving systems of linear equations arising from optimization problems. SLSQP uses an in-place version of QR factorization, which requires storing intermediate results for the next steps of the algorithm. Although DAG-based approaches for QR factorization are prevalent in the literature, they often lack control over the intermediate kernel results, providing only the final output matrices Q and R. This limitation is particularly challenging in SLSQP, where intermediate results of QR factorization are crucial for back-substitution logic at each iteration. Our work introduces novel scheduling techniques using a two-queue approach to execute the QR factorization kernel effectively. This approach, implemented in high-level C++ programming language, facilitates compiler optimizations and allows storing intermediate results required by back-substitution logic. Empirical evaluations demonstrate substantial performance gains, including a 10x improvement over the sequential QR version of the SLSQP algorithm.

\keywords{Non Linear Programming \and Parallel Computing \and DAG Sch-eduling.}
\end{abstract}

\section{Introduction}
\label{sec:intro}
In modern engineering, the demand for efficient optimization techniques is critical across disciplines such as structural engineering, material sciences, and molecular dynamics. Nonlinear programming (NLP) has emerged as a fundamental tool for addressing complex design challenges, as these problems involve intricate, nonlinear relationships that govern system performance and reliability. In structural engineering, NLP facilitates the optimization of large-scale structures, while in material sciences, it enables the development of novel materials with tailored properties. Similarly, molecular dynamics relies on nonlinear equations to model particle interactions, requiring advanced optimization techniques to accurately capture complex behaviors. As the complexity and dimensionality of design spaces continue to expand, advanced computational methods are essential for achieving efficient and scalable solutions.

\noindent
\textbf{SLSQP:} Sequential Least Squares Quadratic Programming \cite{SLSQP} is a well established algorithm for solving NLP problems involving constrained, smoo-th, and differentiable functions. At each iteration, SLSQP constructs a quadratic approximation of the nonlinear objective function while employing a linearization of the constraints, resulting in a quadratic programming (QP) subproblem that determines the search direction for updating decision variables. Despite its effectiveness, the sequential execution of core linear algebra operations poses computational challenges, particularly in high-dimensional optimization problems. As problem dimensionality increases, these sequential computations introduce performance bottlenecks, necessitating the exploration of more efficient and scalable approaches\cite{gill2015performance}.


\noindent
\textbf{QR Factorization:} Enhancing algorithm efficiency is key for progress in optimi-zation driven fields. The QR Factorization is a mathematical technique used to decompose a matrix \(A\) into an orthogonal matrix \(Q\) and an upper triangular matrix \(R\). Methods like Householder transformations achieve this by iteratively applying a sequence of reflection matrices \(H_k\) to \(A\). Each \(H_k\) is constructed from the current state of the \(k\)-th column. The \emph{critical intermediate results} of this process are the components defining these Householder reflectors. The final \(R\) matrix resides in the upper triangle, while \(Q\) is implicitly represented as the product of the \(H_k\) transformations (\(Q = H_1 H_2 \dots H_m\)). In algorithms like SLSQP, which iteratively solve systems of equations or least-squares problems arising from Quadratic Programming (QP) subproblems, these stored intermediate results are paramount. They allow for the efficient application of \(Q\) or \(Q^T\) to various matrices and vectors without explicitly forming the (potentially dense) matrix \(Q\). This repeated application is fundamental to updating solutions and Lagrange multipliers within SLSQP. QR Factorization is a critical sub-routine of SLSQP that is invoked multiple times, making the management and parallel computation of these intermediate results crucial for overall algorithmic performance.

This study presents parallel techniques for QR factorization that harness the advantages of concurrent task execution. By decomposing QR factorization into smaller, independent tasks suitable for parallel processing, an asynchronous Directed Acyclic Task Graph (DATG) scheduling mechanism is employed to optimize execution while preserving task dependencies.

Empirical evaluations demonstrate substantial performance gains in solving large-scale NLP problems using the SLSQP algorithm. The results underscore the transformative impact of parallel computing techniques on QR factorization, reinforcing the necessity for high-performance numerical methods within open-source optimization frameworks.

\noindent \textbf{Our Contributions:} The key contributions of this work are as follows.
\begin{itemize}
	\item  Developed a dynamic algorithm that schedules QR factorization into smaller tasks using asynchronous DATG scheduling\hyperref[alg:thdwork]{Alg. 5}.
	\item Integrated the optimized parallel QR factorization method into the SLSQP implementation of the open-source NLOPT library. 
	\item Comprehensive evaluations demonstrated a 10x improvement of the parallel QR technique over the sequential QR version of the SLSQP algorithm. 
\end{itemize}

\section{Background}
\label{sec:bkgd}
\subsection{SLSQP $\rightarrow$ Descent Direction Computation $\rightarrow$ QR}
The Sequential Least Squares Programming (SLSQP) problem is formulated as a constrained optimization problem, typically expressed in the following mathematical form:

\begin{equation}
    \min_{x} f(x),
\end{equation}
\begin{equation}
    \text{subject to:} \quad h(x) = 0, \quad h: \mathbb{R}^n \to \mathbb{R}^m,
\end{equation}
\begin{equation}
    g(x) \leq 0, \quad g: \mathbb{R}^n \to \mathbb{R}^p.
\end{equation}

SLSQP is an efficient algorithm for solving NLP problems subject to equality and inequality constraints. It seeks to find the minimum of a non-linear objective function while ensuring the satisfaction of constraints. The core of SLSQP involves iteratively approximating the solution using a line search approach to compute the descent direction.

SLSQP determines the descent direction by solving a QP sub-problem at each iteration. This QP is derived from the first and second derivatives, gradients and Hessians, of the Lagrangian function associated with the objective function and constraints. QR factorization plays a pivotal role in this process by providing an efficient method to solve the system of linear equations arising from the Karush-Kuhn-Tucker (KKT) conditions\cite{cfc5dc07425343f08f3c8ee5ae8f7ddc}.


Using QR factorization, SLSQP ensures numerical stability and efficiency in solving the KKT system, thereby accelerating the computation of descent directions in constrained optimization problems.

\subsection{QR Factorization using Householder Transformations}

Given a matrix \( A \) of size \( m \times n \), the goal is to compute an orthogonal matrix \( Q \) and an upper triangular matrix \( R \) such that: $A = QR$. The Householder algorithm achieves this by iteratively constructing and applying reflection vectors to transform A into Q and R, either in-place or out-of-place. Considering the importance of \qrf in \slsqp, we next consider efficient ways to parallelize \qrf. 


\begin{algorithm}
	\caption{QR Factorization using Householder Reflections}
	\label{alg:householder_qr}
	\begin{algorithmic}[1]
		\State \textbf{Initialize:} Set \( Q = I \) (identity matrix) and \( R = A \).
		\For{each column index \( j = 1 \) to \( \min(m, n) \)}
		\State Extract column vector \( x = R[j:m, j] \).
		\State Compute \( \alpha = -x_1 \cdot \|x\| \).
		\State Set \( u = x + \alpha e_1 \), where \( e_1 \) is the first standard basis vector.
		\State Normalize \( u = u / \|u\| \).
		\State Compute Householder matrix \( W = I - 2 \frac{uu^T}{\|u\|^2} \).
		\State Apply transformation: \( R \gets W R \).
		\State Update \( Q \gets Q W^T \).
		\EndFor\\
		\Return \( Q, R \) satisfying \( A = QR \).
	\end{algorithmic}
\end{algorithm}

\section{Our Methodology: Efficient Parallel QR Factorization}
\label{sec:meth}
Algorithm \hyperref[alg:householder_qr]{1} outlines QR decomposition's mathematical logic via Householder reflections. This formulation is further expressed in Algorithm \hyperref[alg:alg_inplace_qr]{2}, representing the standard computational structure frequently employed in linear algebra kernels such as QR factorization, Cholesky decomposition, and LU decomposition. The SLSQP algorithm from the NLOPT library uses an in-place QR factorization technique based on Householder transformations.

Algorithm \ref{alg:alg_inplace_qr} represents the in-place transformation of matrix A into the upper triangular matrix R. The algorithm relies on two key computational kernels: 1) \texttt{update\_pivot\_row} and 2) \texttt{update\_trailing\_non\_pivot\_row}. Both kernels operate at the row level, updating individual elements with a linear time complexity, each involving a fixed number of arithmetic operations.
\begin{algorithm}
	\caption{Transform matrix $A$ to upper-triangular form.} \label{alg:alg_inplace_qr}
	\begin{algorithmic}[1]
		\State \textbf{Input:} $A$, a $m \times n$ non-singular real matrix.
		\For{$i = 1$ \textbf{to} $m$}
		\State $(up, b) \gets \Call{\texttt{update\_pivot\_row}}{A, i}$
		\For{$j = i+1$ to $n$}
		\State \Call{\texttt{update\_trailing\_non\_pivot\_row}}{A, i, j, up}
		\EndFor
		\EndFor
		\State \textbf{Output:} Matrix $A$ in upper-triangular form.
	\end{algorithmic}
\end{algorithm}

In Algorithm \hyperref[alg:alg_inplace_qr]{2}, for a given value of $i$, $1\leq i\leq m$, all tasks $T_{i,*}$ represent computations at the $i^{th}$ iteration of the outer loop. The pivot update calculation of the $i^{th}$ row is performed first (task $T_{i,i}$), representing a call to the kernel \texttt{update\_pivot\_row}. Then all the rows of the entire trailing sub-matrix, with $j > i$, are updated (task $T_{i,j}$), by a  call to \texttt{update\_trailing\_non\_pivot\_row}. 



To model the dependency constraints between tasks, we construct a directed acyclic graph (DAG) \hyperref[fig:task_graph]{Fig. 1(b)}, where the vertices represent tasks, and the edges encode dependencies. An edge $e: T \rightarrow T '$ indicates that $ T '$ can start only after $ T $ is completed, regardless of the availability of resources. Each $T_{i,j}$ task in \hyperref[fig:task_graph]{Fig. 1(b)} represents a call to the kernel \texttt{update\_trailing\_non\_pivot\_row}, which can be executed in parallel once its parent tasks are complete. Therefore, any task $T_{i,j}$ always has two parent nodes $T_{i,i}$ and $T_{i-1,j}$ (except for all tasks $T_{i,j}$ in the first level of \hyperref[fig:task_graph]{Fig. 1(b)} where it has only one parent) whose execution must be completed before the execution of task $T_{i,j}$ can begin. The dependency of $T_{i,j}$ on $T_{i,i}$ denotes the dependency within a single iteration according to Algorithm \hyperref[alg:alg_inplace_qr]{2} where the pivot row update needs to be completed first before proceeding with the row updates of the non-pivot rows. However, the dependency of $T_{i,j}$ on $T_{i-i,j}$ denotes the dependency across iterations where a row in the given input matrix can only be updated in the current iteration if it had been successfully updated by a call to the kernel \texttt{update\_trailing\_non\_pivot\_row} in the previous iteration.

\tikzstyle{startstop} = [rectangle, rounded corners, minimum width=3cm, minimum height=1cm, text centered, draw=black]
\tikzstyle{process} = [rectangle, minimum width=3cm, minimum height=1cm, text centered, draw=black]
\tikzstyle{decision} = [diamond, minimum width=3cm, minimum height=0.8cm, text centered, draw=black]
\tikzstyle{arrow} = [thick,->,>=stealth]
\begin{figure}
	\begin{subfigure}{0.5\textwidth}
		\centering
		\begin{tikzpicture}[node distance=1.8cm, scale=0.7, transform shape, xshift=0cm]
			
			\node (root) [startstop, fill=red!30] {Root Node};
			\node (mq) [process, below of=root, fill=blue!30] {Main Queue};
			\node (executor) [process, below of=mq, fill=green!30] {Executor};
			\node (dep) [decision, right of=executor, xshift=3cm, fill=yellow!30] {Dep Satisfied? };
			\node (wq) [process, below of=dep, yshift=-2cm, fill=purple!30] {Wait Queue};
			\node (terminate) [startstop, below of=executor, yshift=-2cm, fill=orange!30] {Terminate};
			
			\draw [arrow] (root) -- node[right] {Enqueue} (mq);
			\draw [arrow] (mq) -- node[right] {Dequeue} (executor);
			\draw [arrow] (executor.east) -- node[above] {If children} (dep.west);
			\draw [arrow] (executor.south) -- node[right] {If no children} (terminate.north);
			
			\draw [arrow] (dep.east) -- node[above] {No} ++(0.5,0) --node[midway,left] {Enqueue} ++(0,-3.8) -- (wq.east);
			
			\draw [arrow] (wq.west) -- ++(-0.5,0) -- ++(-0.3, 0) -- node[midway,right] {Dequeue} ++(0.0, 2) -- ++(2.3,0) -- (dep.south);
			
			\draw [arrow] (dep.north) |- node[right] {Yes} (mq.east) node[right, above]{\quad\quad\quad\quad\quad\quad\quad\quad Enqueue};
			
		\end{tikzpicture}
		\caption{Flowchart for Task Execution}
		\label{fig:flowchart}
	\end{subfigure}
	\hfill
	\begin{subfigure}{0.45\textwidth}
		\begin{tikzpicture}[node distance=1cm and 0.5cm, scale=0.55, transform shape]
			\node[draw, circle, fill=lightblue] (T11) at (0,0) {$T_{1,1}$};
			\node[draw, circle, fill=lightblue] (T12) at (1.4,-1.2) {$T_{1,2}$};
			\node[draw, circle] (T13) at (2.6,-1.2) {$T_{1,3}$};
			\node[draw, circle] (T14) at (3.8,-1.2) {$T_{1,4}$};
			\node[draw, circle] (T15) at (4.9,-1.2) {$T_{1,5}$};
			
			\node[draw, circle, fill=lightblue] (T22) at (1.4,-2.5) {$T_{2,2}$};
			\node[draw, circle, fill=lightblue] (T23) at (2.6,-3.5) {$T_{2,3}$};
			\node[draw, circle] (T24) at (3.8,-3.5) {$T_{2,4}$};
			\node[draw, circle] (T25) at (4.9,-3.5) {$T_{2,5}$};
			
			\node[draw, circle, fill=lightblue] (T33) at (2.6,-4.8) {$T_{3,3}$};
			\node[draw, circle, fill=lightblue] (T34) at (3.8,-5.8) {$T_{3,4}$};
			\node[draw, circle] (T35) at (4.9,-5.8) {$T_{3,5}$};
			
			\node[draw, circle, fill=lightblue] (T44) at (3.8,-7.1) {$T_{4,4}$};
			\node[draw, circle, fill=lightblue] (T45) at (4.9,-8.1) {$T_{4,5}$};
			
			\node[draw, circle, fill=lightblue] (T55) at (4.9,-9.7) {$T_{5,5}$};
			
			\draw[->] (T11) -- (T12);
			\draw[->] (T12) -- (T22);
			\draw[->] (T13) -- (T23);
			\draw[->] (T14) -- (T24);
			\draw[->] (T15) -- (T25);
			\draw[->, bend left] (T11) to (T13);
			\draw[->, bend left] (T11) to (T14);
			\draw[->, bend left] (T11) to (T15);
			
			\draw[->] (T22) -- (T23);
			\draw[->] (T24) -- (T34);
			\draw[->] (T25) -- (T35);
			\draw[->, bend left] (T22) to (T24);
			\draw[->, bend left] (T22) to (T25);
			
			\draw[->] (T23) -- (T33);
			\draw[->, bend left] (T33) to (T34);
			\draw[->, bend left] (T33) to (T35);
			
			\draw[->] (T34) -- (T44);
			\draw[->] (T35) -- (T45);
			\draw[->, bend left] (T44) to (T45);
			
			\draw[->] (T45) -- (T55);
		\end{tikzpicture}
            \caption{TaskGraph for Triangular System}
		\label{fig:task_graph}
	\end{subfigure}
	
	\caption{Task Execution Flowchart and Task Graph}
	\label{fig:comparison}
\end{figure}
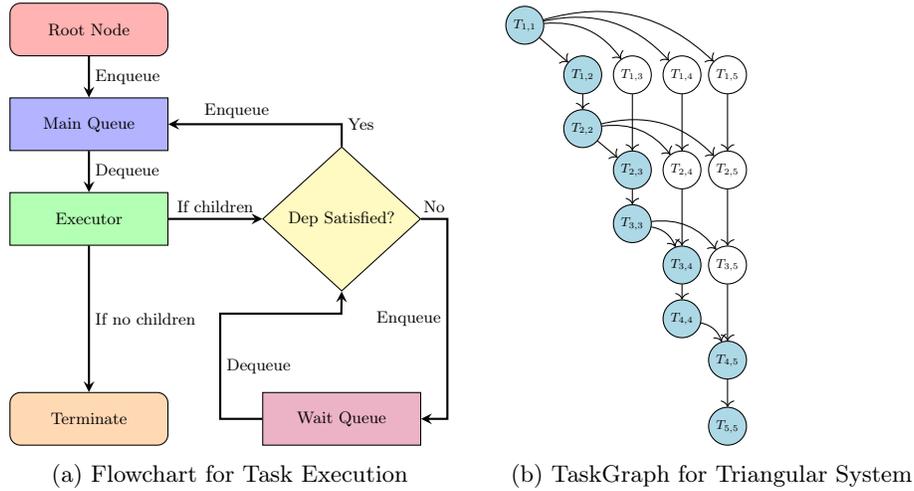

\subsection{Optimizing Thread Workload for Parallel Task Execution}
Even though all the tasks $T_{i,j}$ describe the availability of parallel tasks for the threads, however, the call to the kernel \texttt{update\_trailing\_non\_pivot\_row} only updates a single non-pivot row. We, therefore, want a mechanism to control the amount of work per thread. We introduce two control parameters $\alpha$ and $\beta$ and design two new kernels \hyperref[alg:complete_task1]{Task 1} and \hyperref[alg:complete_task2]{Task 2}, which allows us to increase/decrease the amount of work available per thread by coalescing smaller tasks into larger chunks. The parameter $\beta$ determines how many non-pivot rows each thread updates simultaneously using the \texttt{update\_trailing\_non\_pivot\_row} kernel, rather than processing one row at a time. The parameter $\alpha$ controls the number of iterations in which these $\beta$ rows are updated once the pivot computations for the $\alpha$ rows are complete. Accordingly, \hyperref[alg:complete_task1]{Task 1} performs $\alpha$ pivot computations, enabling efficient batched updates of non-pivot rows in chunks of $\beta$ (\hyperref[alg:complete_task2]{Task 2}) over $\alpha$ iterations.

\subsection{DAG Scheduling using Barriers}
Given the task graph $G = (V, E)$ in \hyperref[fig:task_graph]{Fig. 1(b)}, where $V$ represents the set of nodes and $E$ represents the set of directed edges.  A directed edge $(i, j)$ between two task nodes $t_i$ and $t_j$ indicates that $t_i$ must be completed before $t_j$ can commence. A straightforward approach to schedule the DAG in \hyperref[fig:task_graph]{Fig. 1(b)} across multiple processors while preserving the dependencies is to use \textbf{barriers}—a synchronization mechanism that ensures all threads reach a specific point before proceeding further.  In \hyperref[fig:task_graph]{Fig. 1(b)}, each task \(T_{i,j}\) depends on both \(T_{i,i}\) and \(T_{i-1,j}\), necessitating two synchronization points:  
\begin{algorithm}
	\caption{Function: Task 1}\label{alg:complete_task1}
	\begin{algorithmic}[1]
		\State \textbf{Input:} 
		\begin{itemize}
                \item Matrix \texttt{mat} of size $m\times n$, $pivot\_start$, $row\_chunk\_start$.
		\end{itemize}
		\State \textbf{Global:} \texttt{global\_up\_array}, \texttt{global\_b\_array}, \texttt{$\alpha$}, \texttt{$\beta$}.
		\State $pivot\_end \gets pivot\_start + \alpha $
		\State $row\_chunk\_end \gets row\_chunk\_start + \beta$
		\For{$lpivot = pivot\_start$ \textbf{to} $pivot\_end-1$}
		\State $(up,\, b) \gets \Call{\texttt{update\_pivot\_row}}{mat,\, n,\, lpivot}$
		\If{$up$ or $b$ is \textbf{undefined}}
		\State \textbf{continue}
		\EndIf
		\State \texttt{global\_up\_array}[lpivot] $\gets up$
		\State \texttt{global\_b\_array}[lpivot] $\gets b$
		\For{$j = lpivot+1$ \textbf{to} $row\_chunk\_end-1$}
		\State \Call{\texttt{update\_trailing\_non\_pivot\_row}}{mat,\, n,\, lpivot,\, j,\, up,\, b}
		\EndFor
		\EndFor
		\State \textbf{Output:} Updated matrix \texttt{mat}.
	\end{algorithmic}
\end{algorithm}

\begin{itemize}
    \item A barrier after \(T_{i,i}\) ensures that the \texttt{update\_pivot\_row} kernel completes before executing parallel tasks \(T_{i,j}\).  
    \item A second barrier at the end of each iteration ensures that all tasks \(T_{i,j}\) complete before proceeding to the next iteration, as \(T_{i+1,j}\) depends on \(T_{i,j}\).  
\end{itemize}  

However, barriers impose a rigid execution order, limiting parallel efficiency. For instance, if \(T_{i,j}\) belonging to a critical path completes, the next \(T_{i+1,i+1}\) could begin execution immediately, which after completion can unleash more parallel tasks from the next level. Yet, due to barriers, all threads must wait for the slowest task to complete, even when additional work is available. 
\begin{algorithm}
	\caption{Function: Task 2}
	\label{alg:complete_task2}
	\begin{algorithmic}[1]
		\State \textbf{Input:} 
		\begin{itemize}
			\item Matrix \texttt{mat} of size $m\times n$, $pivot\_start$, $row\_chunk\_start$.
		\end{itemize}
		\State \textbf{Global:} \texttt{global\_up\_array}, \texttt{global\_b\_array}, \texttt{$\alpha$}, \texttt{$\beta$}.
		\State $pivot\_end \gets pivot\_start + \alpha $
		\State $row\_chunk\_end \gets row\_chunk\_start + \beta$
		\For{$lpivot = pivot\_start$ \textbf{to} $pivot\_end-1$}
		\State \(up \gets global\_up\_array[lpivot]\)
		\State \(b \gets global\_b\_array[lpivot]\)
		\For{$j = lpivot+1$ \textbf{to} $row\_chunk\_end-1$}
		\State \Call{\texttt{update\_trailing\_non\_pivot\_row}}{mat,\, n,\, lpivot,\, j,\, up,\, b}
		\EndFor
		\EndFor
		\State \textbf{Output:} Updated matrix \texttt{mat}.
	\end{algorithmic}
\end{algorithm}
\subsection{DAG Scheduling using LockFree Queues}
\label{sec:proposed_alg}
A key observation from \hyperref[fig:task_graph]{Fig. 1(b)} is that traversing the critical path (highlighted nodes) allows more tasks \(T_{i,j}\) to be executed in parallel across different levels. This approach reduces synchronization overhead by requiring only a single dependency check, specifically in \(T_{i-1,j}\). Based on this insight, we propose a dual-queue scheduling mechanism for DAG execution: one queue handles the parallel generation of tasks, while the other ensures that dependencies are satisfied before execution.

As illustrated in \hyperref[fig:flowchart]{Fig. 1(a)}, our approach leverages two centralized, lock-free global queues—\texttt{main\_queue} and \texttt{wait\_queue}—which are shared among all threads for task scheduling. The main queue contains tasks that any available thread can immediately execute. When a critical path task \(T_{i,i}\) is completed, it loads its child tasks \(T_{i,j}\) into the \texttt{main\_queue} after verifying the completion of their parent \(T_{i-1,j}\). If the parent has already completed the task, the child task is immediately available for execution. Otherwise, the task is placed in the wait queue, allowing threads to continue executing readily available tasks from the main queue instead of waiting for the pending parent task to complete. This strategy, as depicted in \hyperref[alg:thdwork]{Algorithm 5} ensures that threads prioritize active execution over spinning or waiting for dependencies to resolve, thereby improving overall responsiveness. Upon completing a task from the main queue, a thread checks the wait queue for deferred tasks. If a task’s parent has completed, it is moved to the main queue for immediate execution. Otherwise, it is enqueued back into the wait queue for re-evaluation in subsequent iterations.

\subsection{DAG scheduling using Priority queues}

In Baskaran's work \cite{baskaran2009compiler}, each vertex in the DAG is associated with two metrics: top level (\texttt{topL}) and bottom level (\texttt{bottomL}). For any vertex \(v\) in DAG \(G\), the \texttt{topL(v)} is defined as the longest length of the path from the root node to the vertex \(v\), excluding \(v\). 
\begin{algorithm}
	\caption{Thread Work}\label{alg:thdwork}
	\begin{algorithmic}[1]
		\State \textbf{Global:} lockfree \texttt{main\_queue}, \texttt{wait\_queue}; dependency\_table \texttt{tb}
		\While{\textbf{true}}
		\If{\texttt{main\_queue} $\neq \emptyset$}
		\State curr\_task $\gets$ \texttt{main\_queue}.pop()
		\If{curr\_task.type = 1}
		\State Task1(curr\_task.params)
		\State \texttt{tb}[curr\_task] = True
		\For{child $\in$ curr\_task.children}
		\If{$\forall p \in$ child.parent, \texttt{tb}[p] = \texttt{True}}
		\State \texttt{main\_queue}.push(child)
		\Else
		\State \texttt{wait\_queue}.push(child)
		\EndIf
		\EndFor
		\ElsIf{curr\_task.type = 2}
		\State Task2(curr\_task.params)
		\State \texttt{tb}[curr\_task] = True
		\If{curr\_task $\in$ CriticalPath}
		\State task1 = curr\_task.children[0]
		\State \texttt{main\_queue}.push(task1)
		\EndIf
		\EndIf
		\EndIf
		\If{\texttt{wait\_queue} $\neq \emptyset$}
		\State old\_task $\gets$ \texttt{wait\_queue}.pop()
		\If{$\forall p \in$ old\_task.parent, \texttt{tb}[p] = \texttt{True}}
		\State \texttt{main\_queue}.push(old\_task)
		\Else
		\State \texttt{wait\_queue}.push(old\_task)
		\EndIf
		\EndIf
		\If{$\exists$ \texttt{task} $\in$ \texttt{tb}, \texttt{tb}[\texttt{task}] = \texttt{False}}
		\State \textbf{continue}
		\Else
		\State \textbf{break}
		\EndIf
		\EndWhile
	\end{algorithmic}
\end{algorithm}

Similarly, \texttt{bottomL(v)} is defined as the length of the longest path from \(v\) to the leaf node (vertex with no children). The tasks are prioritized based on the sum of \texttt{topL(v)} and \texttt{bottomL(v)} or just the \texttt{bottomL(v)}. Nodes that are part of the critical path will have higher priority, and as we move away from the critical path, the priority value of the nodes decreases. We use this technique to assign priority values to each node in the task graph, ensuring that critical-path tasks are executed earlier to accelerate the release of dependent parallel tasks.

Our proposed approach in Section \ref{sec:proposed_alg} employs standard lock-free queues that execute DAG nodes without prioritization. Replacing them with global lock-free \textbf{priority queues} allows nodes to be ordered based on predefined criteria. This prioritization ensures that critical-path nodes execute earlier, thereby accelerating the release of dependent parallel tasks. However, maintaining priority order introduces overhead from rebalancing the data structure, which can degrade performance for large queues. Our implementation utilizes Intel TBB \textbf{concurrent priority queues} and \textbf{concurrent queues} to optimize task scheduling.

\section{Experimental Results}
\label{sec:expts}

\pgfplotsset{compat=1.17}

\definecolor{darkgreen}{rgb}{0,0.5,0}

\pgfplotsset{
    myAxisStyle/.style={
        tick label style={font=\scriptsize},
        label style={font=\small},
        title style={at={(0.5,1.1)}, anchor=south, font=\small},
        grid=both,
        width=6.5cm,
        height=5cm
    },
    myBarStyle/.style={
        tick label style={font=\scriptsize},
        label style={font=\small},
        legend style={nodes={scale=0.75, transform shape}},
        width=12.5cm,
        height=5cm
    }
}

\begin{figure}
    \centering
        \begin{subfigure}{0.5\linewidth}
        \centering
        \includegraphics[width=\linewidth]{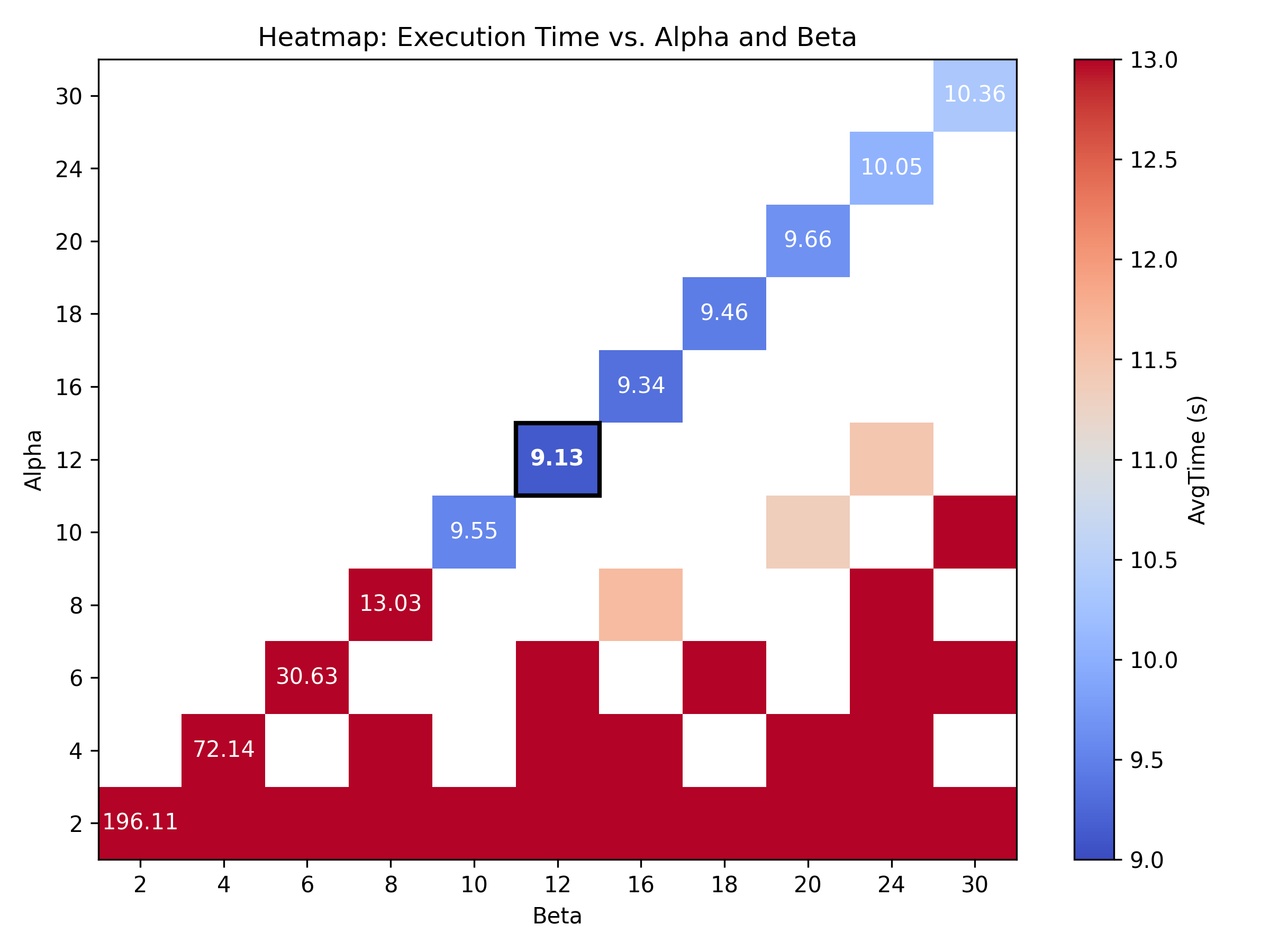}
        \caption{Without Priority}
        \label{fig:exp1_heatmap_a}
    \end{subfigure}\hfill
    \begin{subfigure}{0.5\linewidth}
        \centering
        \includegraphics[width=\linewidth]{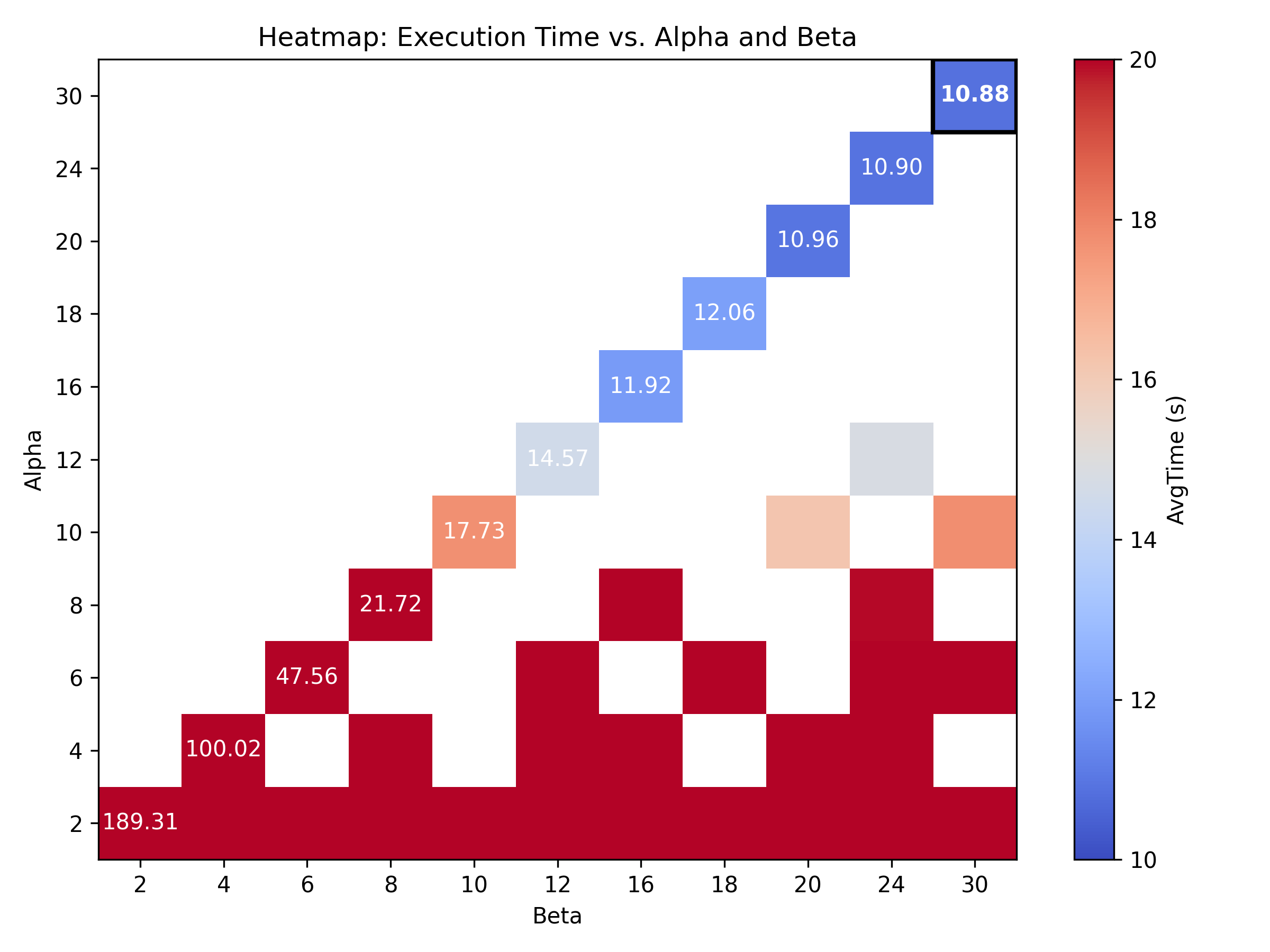}
        \caption{With Priority}
        \label{fig:exp1_heatmap_b}
    \end{subfigure}
    \caption{Heatmap views for the parameter sweep.}
    \label{fig:exp1_heatmap}
\end{figure}

\subsection{Parameter Tuning for Parallel QR Factorization}\label{exp:exp1}
\paragraph{(a) Heatmap Analysis:} 
In this experiment, an exhaustive sweep over the parameters $\alpha$ and $\beta$ (ranging from 2 to 32) was conducted on a fixed matrix size of $10800 \times 10800$ using 26 threads. The primary objective was to minimize execution time. Figure~\ref{fig:exp1_heatmap} presents heatmap visualizations for two cases: without priority scheduling and with priority scheduling. The results reveal that the optimal configuration occurs when $\alpha$ equals $\beta$, with $\alpha=\beta=12$ in the absence of priority scheduling and $\alpha=\beta=30$ under priority-based scheduling.

\begin{figure}
    \centering
    \begin{subfigure}{\linewidth}
        \centering
		\begin{tikzpicture}
			\begin{axis}[
				ybar,
				bar width=7pt,
				symbolic x coords={300,600,1200,2400,3600,4800,6000,7200,8400,9600,10800,12000},
				xtick=data,
				xlabel={\scriptsize	 Matrix Size},
				ylabel={\scriptsize	 Optimal $\alpha$ and $\beta$},
				myBarStyle,
				tick label style={font=\scriptsize},
				label style={font=\scriptsize},
				legend style={font=\scriptsize	, nodes={scale=0.8, transform shape}, legend columns=-1, legend pos=north east},
				ymin=0, ymax=30,
				enlarge x limits=0.1,
				nodes near coords,
				every node near coord/.append style={font=\footnotesize}
				]
				\draw[draw=none,fill=yellow, fill opacity=0.5] 
				(axis cs:300,10) rectangle (axis cs:12000,20);
				
				\addplot coordinates {(1200,16) (2400,24) (3600,24) (4800,20) (6000,20) (7200,18) (8400,16) (9600,16) (10800,12)};
				\addplot coordinates {(1200,16) (2400,16) (3600,20) (4800,20) (6000,20) (7200,20) (8400,20) (9600,16) (10800,18)};
				\addplot coordinates {(1200,12) (2400,12) (3600,12) (4800,16) (6000,15) (7200,15) (8400,12) (9600,12) (10800,12)};
				\legend{26 Threads, 52 Threads, 104 Threads}
			\end{axis}
		\end{tikzpicture}
        \caption{Without Priority}
        \label{fig:exp1_bar_a}
    \end{subfigure}

    \begin{subfigure}{\linewidth}
        \centering
		\begin{tikzpicture}
			\begin{axis}[
				ybar,
				bar width=7pt,
				symbolic x coords={300,600,1200,2400,3600,4800,6000,7200,8400,9600,10800,12000},
				xtick=data,
				xlabel={\scriptsize	 Matrix Size},
				ylabel={\scriptsize	 Optimal $\alpha$ and $\beta$},
				myBarStyle,
				tick label style={font=\scriptsize	},
				label style={font=\scriptsize	},
				legend style={font=\scriptsize	, nodes={scale=0.8, transform shape}, legend columns=-1, legend pos=north east},
				ymin=0, ymax=40,
				enlarge x limits=0.15,
				nodes near coords,
				every node near coord/.append style={font=\footnotesize}
				]
				\draw[draw=none,fill=yellow, fill opacity=0.5] 
				(axis cs:300,10) rectangle (axis cs:12000,25);
				
				\addplot coordinates {(1200,20) (2400,32) (3600,30) (4800,24) (6000,24) (7200,24) (8400,14) (9600,30) (10800,30)};
				\addplot coordinates {(1200,12) (2400,16) (3600,24) (4800,24) (6000,24) (7200,20) (8400,24) (9600,24) (10800,20)};
				\addplot coordinates {(1200,15) (2400,15) (3600,20) (4800,15) (6000,15) (7200,15) (8400,12) (9600,12) (10800,10)};
				\legend{26 Threads, 52 Threads, 104 Threads}
			\end{axis}
		\end{tikzpicture}
        \caption{With Priority}
        \label{fig:exp1_bar_b}
    \end{subfigure}
    
    \caption{Bar graphs of best configurations across different matrix sizes.}
    \label{fig:exp1_bar}
\end{figure}
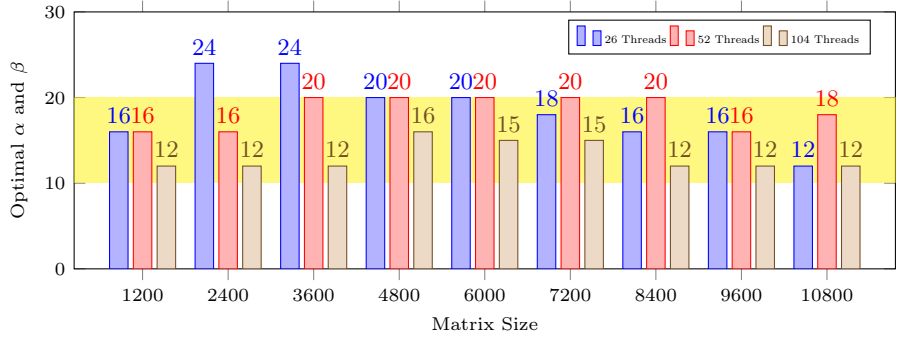
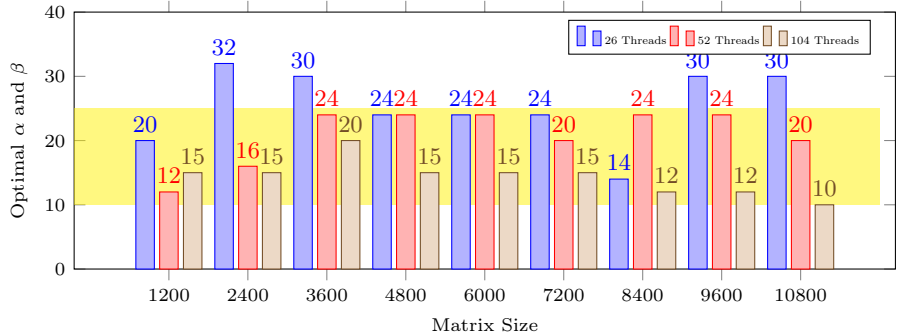

\paragraph{(b) Bar Graph Analysis:} Figure~\ref{fig:exp1_bar} illustrates the evolution of optimal $\alpha=\beta$ settings across various matrix sizes and thread counts (26, 52, and 104 threads), highlighting how computational load influences parameter tuning. The yellow highlighted band indicates the range in which the optimal parameter values consistently fall across different matrix sizes.
A key insight from the results is that, as the number of threads increases, the variability in the optimal $\alpha$–$\beta$ values diminishes. This convergence suggests that, under higher parallelism, finer granularity (i.e., lower $\alpha$ and $\beta$) is preferred to maximize workload distribution and minimize execution time.

\subsection{Scalability Analysis}\label{exp:exp2}
In this experiment, we assesses the scalability of our proposed algorithm on dense square matrices with dimensions ranging from $300 \times 300$ to $10800 \times 10800$. The optimal $\alpha$ and $\beta$ values determined in Experiment~\ref{exp:exp1} were employed consistently. To capture the impact of parallelism, tests were performed using 26 and 52 threads.

To assess the effectiveness of our approach, we compare three different methods: (i) a parallel DAG execution method that employs synchronization barriers and (ii) two variants of our proposed approach—one incorporating a priority-based scheduling mechanism and another without priority. This comparative analysis provides insights into the efficiency and scalability of the proposed methodology under varying computational loads.

Figure~\ref{fig:exp2_a} (26 threads) and Figure~\ref{fig:exp2_b} (52 threads) display the execution time trends as the matrix size increases. The results clearly demonstrate that both variants of our method (with and without priority scheduling) significantly outperform the barrier-based parallel DAG execution. However, the priority-free variant performs slightly better than priority-based due to the overhead introduced by priority queues. The additional overhead stems from the fact that priority queues reorder nodes according to their priority values and require more frequent data structure re-balancing.

\begin{figure}
    \centering
    \begin{subfigure}{0.5\linewidth}
        \centering
        \begin{tikzpicture}
            \begin{axis}[
                xlabel={Matrix Size},
                ylabel={Execution Time (s)},
                xtick={300, 2400, 4800, 7200, 10800},
                xticklabels={300, 2400, 4800, 7200, 10800},
                ytick={0, 10, 20, 30, 40, 50},
                yticklabels={0, 10, 20, 30, 40, 50},
                scaled ticks=false,
                tick label style={/pgf/number format/fixed},
                myAxisStyle,
                legend style={nodes={scale=0.75, transform shape}},
                legend pos=north west
                ]
                \addplot[darkgreen, mark=*] coordinates {
                    (300,0.009) (512,0.019) (600,0.02467) (1024,0.072) (1200,0.095) (2048,0.31433) 
                    (2400,0.472) (3072,0.97033) (3600,1.60833) (4096,2.249) (4800,3.50733) 
                    (6000,7.049) (7200,11.83833) (8192,18.866) (8400,19.69433) (9600,30.26433) 
                    (10800,43.42867)
                };
                \addlegendentry{Barrier};
                
                \addplot[red, mark=square*] coordinates {
                    (300,0.002) (512,0.00433) (600,0.00533) (1024,0.01733) (1200,0.02067) (2048,0.06333) 
                    (2400,0.089) (3072,0.183) (3600,0.289) (4096,0.42933) (4800,0.66933) 
                    (6000,1.35233) (7200,2.41633) (8192,3.72667) (8400,4.063) 
                    (9600,6.28833) (10800,9.13267)
                };
                \addlegendentry{Without Priority};
                
                \addplot[blue, mark=triangle*] coordinates {
                    (300,0.002) (512,0.005) (600,0.00567) (1024,0.01667) (1200,0.021) (2048,0.06133) 
                    (2400,0.08733) (3072,0.18533) (3600,0.30633) (4096,0.44233) (4800,0.73167) 
                    (6000,1.53133) (7200,2.85767) (8192,4.401) (8400,4.67167) 
                    (9600,7.387) (10800,10.87967)
                };
                \addlegendentry{With Priority};
            \end{axis}
        \end{tikzpicture}
        \caption{Exec. Time vs Matrix Size (26 Threads)}
        \label{fig:exp2_a}
    \end{subfigure}\hfill
    \begin{subfigure}{0.5\linewidth}
        \centering
        \begin{tikzpicture}
            \begin{axis}[
                xlabel={Matrix Size},
                ylabel={Execution Time (s)},
                xtick={300, 2400, 4800, 7200, 10800},
                xticklabels={300, 2400, 4800, 7200, 10800},
                ytick={0, 10, 20, 30, 40, 50},
                yticklabels={0, 10, 20, 30, 40, 50},
                scaled ticks=false,
                tick label style={/pgf/number format/fixed},
                myAxisStyle,
                legend style={nodes={scale=0.75, transform shape}},
                legend pos=north west
                ]
                \addplot[darkgreen, mark=*] coordinates {
                    (300,0.01367) (512,0.02867) (600,0.038) (1024,0.08867) (1200,0.12633) 
                    (2048,0.44633) (2400,0.646) (3072,1.20967) (3600,2.06267) 
                    (4096,2.74733) (4800,4.058) (6000,7.4) (7200,13.15033) 
                    (8192,21.07267) (8400,21.81733) (9600,34.544) (10800,53.54133)
                };
                \addlegendentry{Barrier};
                
                \addplot[red, mark=square*] coordinates {
                    (300,0.00267) (512,0.00567) (600,0.00667) (1024,0.01567) (1200,0.01933) 
                    (2048,0.04767) (2400,0.069) (3072,0.13033) (3600,0.195) (4096,0.28833) 
                    (4800,0.454) (6000,1.00233) (7200,1.838) (8192,3.39833) 
                    (8400,3.21767) (9600,4.96967) (10800,7.22533)
                };
                \addlegendentry{Without Priority};
                
                \addplot[blue, mark=triangle*] coordinates {
                    (300,0.00333) (512,0.007) (600,0.00767) (1024,0.01733) (1200,0.019) 
                    (2048,0.04967) (2400,0.066) (3072,0.12633) (3600,0.191) (4096,0.30833) 
                    (4800,0.45567) (6000,0.99867) (7200,2.00967) (8192,4.25967) 
                    (8400,3.464) (9600,5.58033) (10800,9.47167)
                };
                \addlegendentry{With Priority};
            \end{axis}
        \end{tikzpicture}
        \caption{Exec. Time vs Matrix Size (52 Threads)}
        \label{fig:exp2_b}
    \end{subfigure}
    \caption{Scalability comparison of the proposed algorithm for different matrix sizes.}
    \label{fig:exp2}
\end{figure}
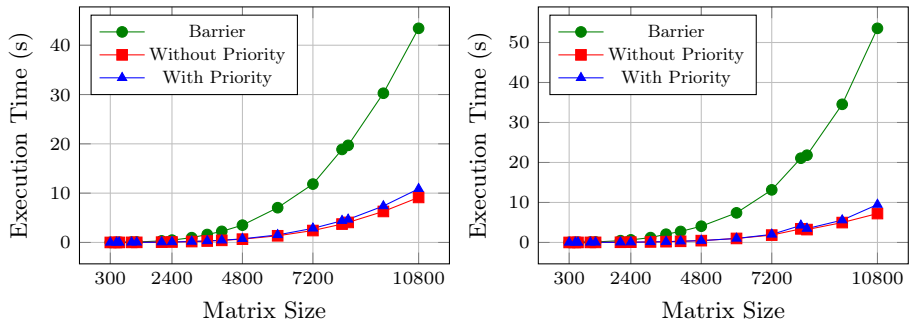

\begin{figure}
    \centering
    \begin{minipage}{0.5\linewidth}
        \centering
        \begin{tikzpicture}
            \begin{axis}[
                xlabel={Threads},
                ylabel={Execution Time (s)},
                myAxisStyle,
                ymode=log,
                log basis y=10,
                xtick={4, 24, 44, 64, 84, 100},
                ytick={10, 100, 1000, 10000, 100000},
                legend style={at={(axis description cs:0.5,0.5)}, anchor=west, nodes={scale=0.57, transform shape}}
                ]
                \addplot[darkgreen, mark=o] coordinates {
                    (4,60.32933) (8,37.08767) (12,29.13767) (16,24.59833)
                    (20,22.211) (24,20.787) (28,20.160) (32,21.67867)
                    (36,23.511) (40,21.980) (44,21.87833) (48,24.22767)
                    (52,27.220) (56,27.414) (60,27.967) (64,28.367)
                    (68,28.91533) (72,29.754) (76,30.85733) (80,32.056)
                    (84,32.97067) (88,34.05467) (92,35.255) (96,36.73167)
                    (100,38.23033)
                };
                \addlegendentry{Barrier};
                
                \addplot[red, mark=square] coordinates {
                    (4,19.068) (8,9.88267) (12,6.753) (16,5.495)
                    (20,4.492) (24,3.996) (28,3.563) (32,3.33167)
                    (36,3.169) (40,3.258) (44,3.030) (48,2.87633)
                    (52,2.87967) (56,2.85433) (60,2.82233) (64,2.883)
                    (68,2.971) (72,3.06733) (76,3.15033) (80,3.26467)
                    (84,3.438) (88,3.65933) (92,3.891) (96,4.26467)
                    (100,4.668)
                };
                \addlegendentry{Without Priority};
                
                \addplot[blue, mark=triangle] coordinates {
                    (4,19.20933) (8,10.104) (12,7.13767) (16,6.22233)
                    (20,5.67667) (24,4.84067) (28,4.59033) (32,4.56433)
                    (36,4.55167) (40,4.48067) (44,4.507) (48,4.11267)
                    (52,3.804) (56,4.03433) (60,3.80433) (64,3.72833)
                    (68,3.518) (72,3.45633) (76,3.65533) (80,3.98367)
                    (84,4.212) (88,4.07933) (92,4.493) (96,4.891)
                    (100,5.32067)
                };
                \addlegendentry{With Priority};
            \end{axis}
        \end{tikzpicture}
        \caption{Throughput Evaluation}
        \label{fig:fig4}
    \end{minipage}\hfill
    \begin{minipage}{0.49\linewidth}
        \centering
        \begin{tikzpicture}
            \begin{axis}[
                xlabel={Degrees of Freedom},
                ylabel={Execution Time (hr)},
                myAxisStyle,
                legend style={nodes={scale=0.75, transform shape}},
                xtick={640,1250,1728,2240,2816},
                ytick={0,2,6,12,18},
                legend pos=north west
                ]
                \addplot[orange, mark=*, thick] coordinates {
                    (640,0.02908)
                    (1250,0.3004)
                    (1728,1.89514)
                    (2240,5.415)
                    (2816,18.91215)
                };
                \addlegendentry{Sequential QR in SLSQP};
                
                \addplot[blue, mark=square*, thick] coordinates {
                    (640,0.001454)
                    (1250,0.02879)
                    (1728,0.16458)
                    (2240,0.456)
                    (2816,1.20716)
                };
                \addlegendentry{Parallel QR in SLSQP};
            \end{axis}
        \end{tikzpicture}
        \caption{SLSQP  Performance}
        \label{fig:fig5}
    \end{minipage}
\end{figure}

\subsection{Throughput Evaluation} \label{exp:exp3}
In this experiment, we evaluate the throughput of various algorithms by incrementally increasing the number of threads (in multiples of 4) for a fixed matrix size of $8192 \times 8192$, while using the optimal $\alpha$ and $\beta$ values from Experiment~\ref{exp:exp1}. Figure~\ref{fig:fig4} plots the execution time (on a logarithmic scale) against thread count for three methods: barrier-based, without priority, and with priority scheduling.

The results align with the previous experiment, confirming that our proposed method (with and without priority) outperforms the parallel DAG execution with barriers. However, the priority-based variant is slightly less efficient due to its additional overhead.


\subsection{Impact of Parallel QR Factorization in SLSQP} \label{exp:exp4}

In this experiment, we examine the influence of integrating our proposed parallel QR approach into the SLSQP framework within the NLOPT library. Our focus is on large-scale boundary value problems involving implicit constitutive relations under both elastic and inelastic responses.\cite{ananthapadmanabhan2023multi}.

To assess performance, we compare a baseline SLSQP implementation equip-ped with sequential QR factorization against an SLSQP version that leverages our parallel QR factorization. We evaluated these approaches over a range of problem sizes defined by degrees of freedom (DOF) equal to 640, 1250, 1728, 2240, and 2816. Here, the term "degrees of freedom" denotes the total number of unknowns, such as nodal displacements, stresses, or auxiliary state variables, emerging from the discretizations of the governing partial differential equations and boundary conditions. Consequently, an increase in DOF results in a proportional increase in the dimension of the system matrix factorized by SLSQP.

Figure~\ref{fig:fig5} shows the execution times (in hours) corresponding to each DOF level for both sequential and parallel QR implementations. Notably, when the DOF reaches 2816, the parallel QR variant completes its tasks in approximately 1.21~hours, whereas the sequential counterpart requires nearly 18.91~hours. This substantial improvement in computational efficiency becomes increasingly pronounced as the DOF grows, reflecting the strong scalability of the parallel approach.

\section{Related Works}
\label{sec:related}

The NLOPT\cite{NLopt} library's SLSQP\cite{SLSQP} algorithm is a recognized open-source, numerically stable solver for nonlinear optimization, valued for its customizability. However, unlike continuously evolving contemporaries such as IPOPT\cite{biegler2009large} (open-source) and SNOPT\cite{gill2005snopt} (commercial), SLSQP has seen limited recent advancements. To address this, Joshy et al. introduced the PySLSQP\cite{joshy2024pyslsqp} package, enhancing SLSQP's utility by bridging Python with the original Fortran code, thereby facilitating easier modification and addressing limitations in current formulations, particularly those in NLOPT.
Concurrently, significant progress has been made in parallel QR decomposition. Buttari et al.\cite{buttari2008parallel} introduced a Parallel Tiled QR factorization method using a DAG for scheduling. Baskaran et al.\cite{baskaran2009compiler} contributed with compiler-assisted dynamic scheduling using lock-based priority queues, and building on this, Roshan et al. (Dathathri et al.)\cite{dathathri2016compiling} developed dynamic scheduling techniques with lock-free priority queues for QR factorization.
This research extends existing approaches by enhancing the NLOPT SLSQP solver through the integration of advanced asynchronous parallel algorithms. Departing from traditional synchronous strategies, this method incorporates novel parallel scheduling techniques inspired by the dynamic and lock-free approaches developed for QR factorization, aiming to improve both performance and scalability. For managing the interdependent tasks within QR decomposition, this work employs a variation of DAG scheduling, drawing upon established efficient DAG scheduling methodologies~\cite{DAG-scheduling}.

\section{Conclusions and Future Work}
\label{sec:conc}
This study integrates a highly parallel QR factorization into NLOPT’s SLSQP routine by decomposing the factorization into numerous independent micro-tasks and coordinating them with an asynchronous, dependency-aware DAG scheduler. The resulting workflow markedly cuts computational overhead and achieves consistent speed-ups across a broad suite of benchmark problems, underscoring the value of fine-grained parallelism in nonlinear constrained optimization.
In future work, we will substitute the classical Householder QR with a tiled implementation \cite{buttari2008parallel}\cite{baskaran2009compiler}\cite{dathathri2016compiling}. Tiling will expose even finer parallel granularity, facilitate NUMA-aware task placement, and enable closed-form selection of tuning parameters $\alpha$ and $\beta$, thereby eliminating costly parameter sweeps. Moreover, tiling naturally reduces queue-transfer contention between wait and execution queues by improving spatial locality and minimizing remote memory traffic.
We further intend to generalize this tiling and scheduling strategy to the full suite of linear-algebra kernels invoked by SLSQP—such as rank-update, triangular solve, and Cholesky-related operations—ultimately delivering a solver that scales robustly on diverse multicore and many-core architectures.

\section{Acknowledgments and Artifact Availability}
We extend our deepest gratitude to Dr. Saravanan, Professor of Civil Engineering, Indian Institute of Technology, Madras, for providing us  with the dataset for the experiment \hyperref[exp:exp4]{4.4}. We also thank the anonymous reviewers for their useful comments. The artifact for this work is available in the Zenodo repository \cite{soumyajit_chatterjee_2025_15602262}.

\small \subsubsection{Disclosure of Interests.} This work is partly funded by CRG 009391 \& 001090, GoI (Government of India).

%
%

\printbibliography
\end{document}